# Resistive Plate Chambers for Imaging Calorimetry – the DHCAL

**José Repond**[a]

[a] *Argonne National Laboratory,
9700 S. Cass Avenue
Argonne, IL 60439, U.S.A.
E-mail*: repond@anl.gov

**On Behalf of the CALICE collaboration**

ABSTRACT: The DHCAL – the Digital Hadron Calorimeter – is a prototype calorimeter based on Resistive Plate Chambers (RPCs). The design emphasizes the imaging capabilities of the detector in an effort to optimize the calorimeter for the application of Particle Flow Algorithms (PFAs) to the reconstruction of hadronic jet energies in a colliding beam environment. The readout of the chambers is segmented into $1 \times 1$ cm$^2$ pads, each read out with a 1-bit (single threshold) resolution. The prototype with approximately 500,000 readout channels underwent extensive testing in both the Fermilab and CERN test beams. This talk presents preliminary findings from the analysis of data collected at the test beams.

KEYWORDS: Calorimetry; Resistive Plate Chambers; Particle Flow Algorithm.

# Contents

# 1. Introduction

### 1.1 Why Use RPCs for Calorimetry

In past colliding beam experiments the energy of jets was measured with the calorimeter (systems) alone. The latter typically featured a small number of towers in which the energy deposited by the particles in a hadronic jet was summed up. In contrast to this approach, Particle Flow Algorithms (PFAs) [1] attempt to measure each particle in a jet individually with the component providing the best momentum/energy resolution. Thus, charged particles are measured with a high precision tracker, photons are measured with an electromagnetic calorimeter and the remaining neutral hadrons are measured with the combination of electromagnetic and hadronic calorimeters. In detailed Monte Carlo studies it was shown [1] that PFAs improve the jet energy resolution roughly by a factor of two compared to the pure calorimetric measurement.

The major challenge of this method is related to the correct identification of energy deposits in the calorimeter as associated with incoming charged tracks (and therefore to be ignored) or with neutral particles (and therefore to be measured). Thus, calorimeters optimized for the application of PFAs will feature an unprecedented segmentation of the readout, leading to large numbers of individual readout channels.

For the low rate environment of a future $e^+e^-$ Linear Collider, Resistive Plate Chambers (RPCs) are seen to be the ideal choice for the active element of the Hadron Calorimeter (HCAL). RPCs are robust, simple in design, cheap, can be made to be very thin (less than 3 mm including the readout), stable in operation, and their readout pads can be segmented in whatever shape deemed necessary [2].

## 1.2 The DHCAL

The DHCAL or Digital Hadron Calorimeter is a large scale prototype of an imaging hadron calorimeter, built as integral part of the program of the international CALICE collaboration [3]. It features up to 54 active layers, each with an area of $96 \times 96$ cm$^2$. To test its calorimetric properties, the DHCAL layers were inserted into a main stack and a tail catcher, which was located downstream of the former. Figure 1 shows a photograph of the main stack containing 39 layers and the tail catcher, as configured for tests in the CERN PS and SPS beams.

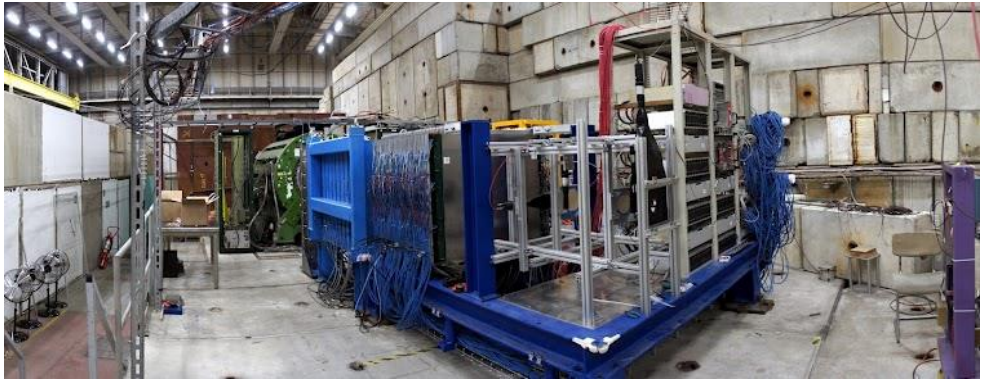

**Figure 1.** Photograph of the DHCAL main stack and tail catcher, as configured for the CERN PS and SPS test beams

The readout of the chambers was segmented into $1 \times 1$ cm$^2$ pads placed directly on the anode resistive plate. The readout boards contain the pads as well as the front-end electronics, based on the DCAL III chip [4]. Each DCAL chip is connected to 64 readout pads. A threshold is applied to the signals from the individual pads. The resulting hits, together with a time stamp with 100 ns resolution, are sent to a back-end readout system located in a VME crate off the detector. The data can be acquired in trigger-less (all hits being recorded) or triggered mode (only hits corresponding to an external trigger are passed on).

In this type of calorimeter the energy of an incident particle is reconstructed to first order as being proportional to the total number of pads with signals above threshold. However, by exploiting the highly imaging capabilities of the device, the raw energy measurement can be refined by identifying electromagnetic sub-showers and applying different weights to their corresponding hits, a technique named software compensation.

## 1.3 DHCAL Construction

The construction of the DHCAL started in fall 2008 and was completed by spring 2011. The assembly of the chambers and the construction of the readout electronics proceeded in parallel. The RPCs were based on the standard 2-glass design and measured $32 \times 96$ cm$^2$. Figure 2 shows a sketch of the cross section of one edge of a DHCAL chamber. The resistive plates were made of soda-lime float glass with a bulk resistivity of $4.7 \cdot 10^{13}$ Ωcm. The side with the thinner glass plate was chosen as the anode and featured the readout board. The outer surfaces of the chambers were sprayed with resistive paint with a surface resistivity of $1 - 8$ MΩ/□. The relatively large surface resistivity was chosen to minimize the average pad multiplicity (see below).

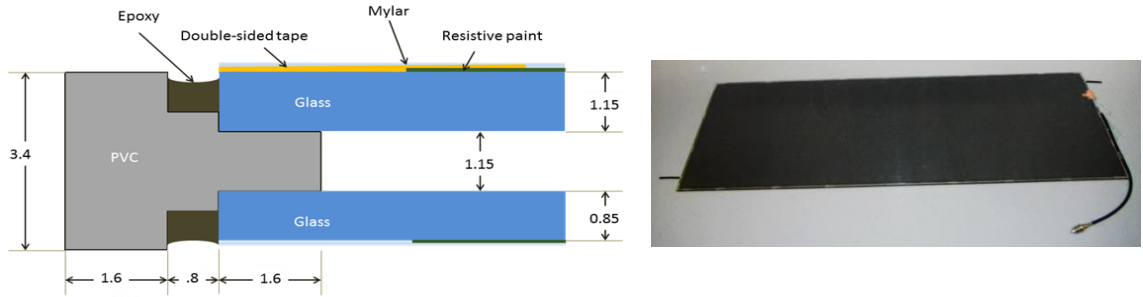

**Figure 2.** Sketch of the cross section of one side of the chamber (left). Photograph of a completed chamber including its high voltage connection (right).

In a so-called engineering run, approximately 10,000 DCAL III chips were produced. Before embarking on the assembly of the readout boards, the chips underwent extensive testing using a fully automated, robotic system. Each readout board measured $32 \times 48$ cm$^2$ and contained 1,536 readout pads, which were connected to 24 DCAL III chips. A data concentrator, located on one edge of the boards, provided the control functions and the link to the back-end of the readout system. A trigger and timing module (TTM), located in the VME crates of the back-end readout system, synchronized the clocks and trigger signals for the entire system. A photograph of a fully assembled readout board is shown in Fig. 3.

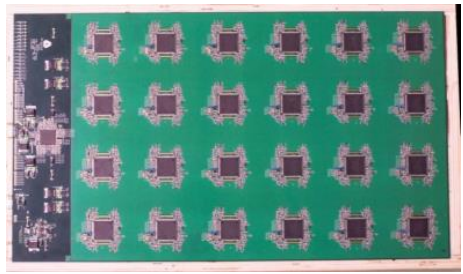

**Figure 3**. Photograph of a DHCAL readout board. The dark area on the left edge of the board corresponds to the data concentrator.

Three chambers and six readout boards were assembled into a detector cassette. For mechanical stability, the cassettes consisted of a copper front plate and a steel back plate, each 2 mm thick. Copper was chosen for the front plate to provide uniform cooling to the DCAL chips, which were in direct contact with the plate.

Extensive tests were performed at every step in the construction and assembly process to assure the quality of the products. As a consequence, after installation and ten minutes after delivery of the first particle beams at the Fermilab FTBF test beam, the DHCAL was able to provide pictures of events showing clean muon tracks traversing the calorimeter.

**1.4 The DHCAL in Test Beams**

Testing in the Fermilab MT6 beam started in October 2010 and lasted until November 2011. Over this period of time, the DHCAL was exposed to beam for a total of about five months. The DHCAL layers were inserted into the CALICE Analog Hadron Calorimeter (AHCAL) structure [5], which featured Steel absorber plates with a thickness of 17.4 mm.

In early 2012 the DHCAL layers were transported to CERN to be tested together with Tungsten absorber plates. The CERN tests included data taking at the Proton-Synchrotron (PS) as well as in the H8 test beam line of the Super-Proton-Synchrotron (SPS). The data taking at CERN took approximately five weeks.

In both locations the main stack was complemented with a tail catcher equipped with Steel plates and located downstream of the main stack. The tail catcher utilized the same DHCAL technology as active elements as did the main stack. Table I summarizes the number of events collected at Fermilab and CERN.

| Test Beam | Muon events | Secondary beam |
|---|---|---|
| Fermilab | 9.4 M | 14.3 M |
| CERN | 4.9 M | 22.1 M |
| Total | 14.3 M | 36.4 M |

**Table I.** Summary of the data collected with the DHCAL at Fermilab and CERN. The secondary beam data includes electrons, muons, pions, and protons, the mixture being dependent on the location, the beam momentum and beam setting.

**2. Results from the Test Beam**

In the following the major results from the various test beam campaigns are briefly summarized. As of today, the data are still being analysed and the present results are considered preliminary. Nevertheless, they play an important role in validating the novel concept of digital hadron calorimetry.

**2.1 Measurements with Muons**

Muons provide an excellent tool to assess the detailed performance of individual calorimeter layers. At Fermilab, for instance, muons were produced with the 32 GeV/c secondary beam impinging on a 3meter long Iron block. The resulting muons have a broad band of momenta, but are still close to minimum ionizing. Figure 4 (left) shows the event display of a muon track through the DHCAL. Notice the absence of accidental hits, as the noise rate in the detector is extremely small.
The performance criteria for the chambers include the MIP detection efficiency, the average pad multiplicity and the product of the two, commonly named the calibration factors. The latter are used to equalize the response of the individual RPCs and, for better illustration, are divided by the calibration factors averaged over the entire calorimeter. The performance criteria of the RPCs, shown in Figure 4 (right), are observed to be fairly uniform, with small variations from layer-to-layer. Larger fluctuations are seen with the average pad multiplicity, compared to the efficiency, which is remarkably uniform. Planned changes in the chamber design, to be implemented in future prototypes, will significantly reduce these fluctuations (see below) and thus will simplify the equalization procedure.

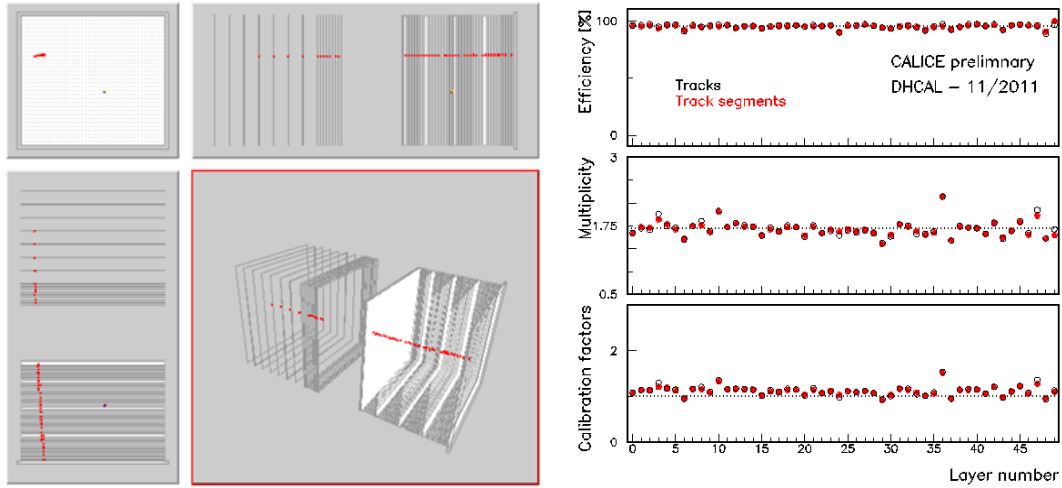

**Figure 4.** Event display of a muon track through the DHCAL (left). Performance criteria as function of layer number as measured at Fermilab with a broadband muon beam (right).

### 2.2 Equalization of the Response of RPCs

The equalization of the response of individual RPCs is necessary to keep the constant term in the resolution function under check. Monte Carlo studies, based on GEANT4 combined with a standalone simulation of the RPC response, were used to explore different equalization schemes. In order to emulate differences in the performance parameters of the RPCs, the threshold applied to the hits in the simulation was varied, where e.g. a higher threshold results in a decrease of the efficiency and average pad multiplicity. Figure 5 shows the effect on varying the threshold for the three different particles of interest (muons, electrons and pions) on the number of hits above threshold. Changes in the threshold are seen to affect the particles differently. This (unexpected) complication led to the need for more sophisticated equalization procedures.

Three different methods have been applied to achieve an equalization of the response:

a) Full calibration: using the muons to equalize the calibration factors;
b) Density weighted calibration: using the density of hits surrounding a given hit to determine a hit's individual weight. This method takes into account the effect that a pad with multiple hits will provide a signal above threshold, even if the chamber's efficiency is below average; and
c) Hybrid calibration: a mixture of the two methods described above.

For a given beam energy, all three methods resulted in a marked decrease of the observed run-to-run variations in the response.

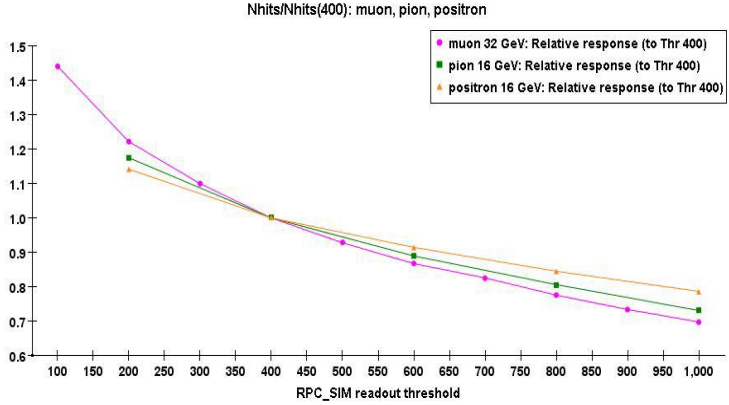

**Figure 5.** Number of hits as function of the applied threshold, normalized by the response at a threshold of 400. The different colors correspond to simulated muons, electrons, and pions.

## 2.3 Measurements with Pions and Fe-absorbers

The response to pions in the energy range from 2 to 60 GeV was measured in the Fermilab test beam. The mean number of hits N as function of beam energy is shown in Figure 6 (left). The results from both the un-calibrated and the density-calibrated sets are shown. The data are fit to a power law function, $N = aE^m$, where the parameter m provides a direct measure of possible saturation effects, with a value of m = 1 corresponding to a perfectly linear behaviour. The fit of the density-calibrated response results in a value of m = 0.98, indicating a 2% saturation of the response. As expected, the calibration procedure results in reduced fluctuations of the individual measurements around the (fitted) power law function.

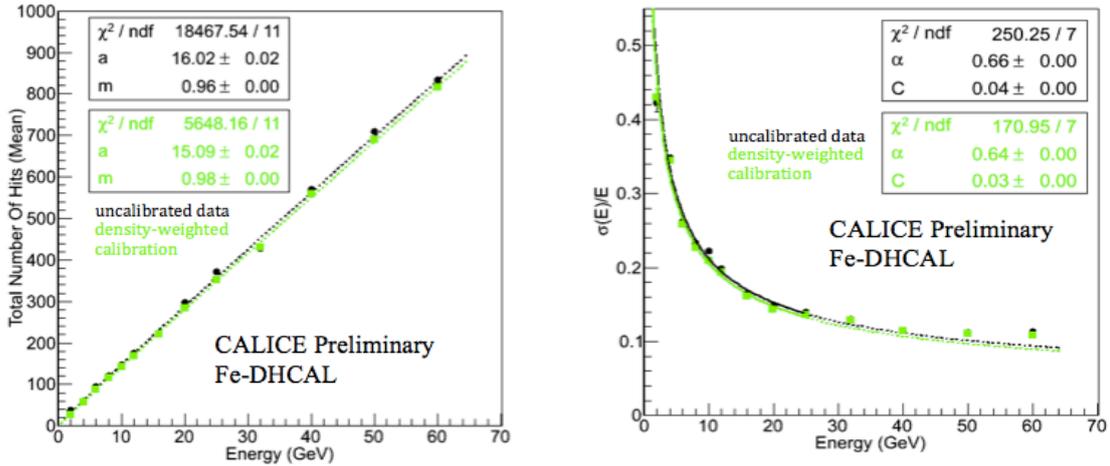

**Figure 6.** Response (mean number of hits) versus beam energy for both un-calibrated (black) and density-calibrated (green) data sets. The lines are the results of fits to a power law, $aE^m$. Energy resolution versus beam energy (right). The solid lines are the results of fits to the sum of a stochastic and a constant term. The fits were performed for points below 30 GeV. The results of the fit are extended to higher energy and shown as dashed curves.

The energy resolution for pions in the range of 2 to 60 GeV is shown in Figure 6 (right). The resolutions below 30 GeV are fitted to the quadratic sum of a constant and stochastic term

$$\frac{\sigma_E}{E} = \frac{\alpha}{\sqrt{E}} \oplus C$$

The stochastic term α = 64% is commensurate with expectations based on Monte Carlo simulations [6]. For higher energies, the resolutions level off and do not improve as fast as expected given the value of the stochastic term.

**2.4 Measurements with W-absorbers**

The data collected at CERN with Tungsten absorber plates covers a wide range of energies from 1 to 300 GeV. The mean number of hits versus beam energy for different particles is shown in Figure 7. A fit of the pion data to the above mentioned power law indicates a significant departure from linearity with m = 0.84. Furthermore, the response to electrons/positrons is seen to be significantly smaller than the response to hadrons of the same energy, with an e/h in the range from 0.5 – 0.9. This makes the W-DHCAL highly over-compensating. In general, due to the increased thickness of the Tungsten absorber plates, compared to the Steel plates, when measured in radiation length, the measured energy resolutions are of the order of 25% worse. Both the use of smaller pads and the application of software compensation techniques are expected to restore the linearity for pions and to improve the resolution and are the subject of future studies.

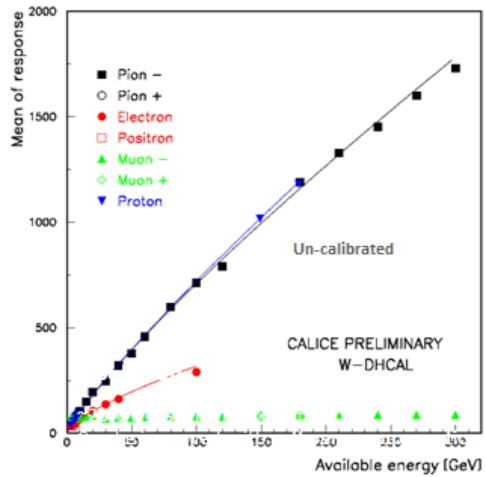

**Figure 7.** Mean number of hits versus beam energy, as measured with the Tungsten stack at CERN. The results are shown for various particles, such as muons, electrons, pions and protons. The results for positive particles are in part covered by the results from negative particles and so not visible.

**3. Operational Issues and Further R&D**

Apart from the benefit of learning about the novel calorimetric aspects, the construction and testing of the DHCAL provided ample experience in operating such a device. Several technical issues surfaced during the prolonged operation in the test beams. Further R&D on both the RPCs and the readout system is being pursued to address these problems.

### 3.1 Operational Issues

In general, the DHCAL behaved very well in the test beam and provided large samples of high quality data. Nevertheless, the following design/operational issues were encountered:

- The efficiency of the chambers was found to drop towards their edges. This loss of efficiency was traced back to a slight increase in the gas gap, in turn due to a slight dimensional irregularity in the extruded channels used as rims. With better rims, this problem is easily resolved.

- Over the extended running periods a number of chambers (about 5%) lost their efficiency, some almost completely. This effect could in part be compensated by an increase in the applied high voltage. The origin of the problem is attributed to a loss of high voltage contact to the resistive paint. It appears that the current at the high voltage leads initiated a chemical reaction which rendered the surface paint non-conductive. In the future, the DHCAL group plans on using resistive Kapton films to provide a resistive surface, which should not be prone to these problems.

- In the last running period, the efficiency in the centre of the chambers was found to have decreased. This effect seems to have originated from a warping of the readout boards over time. The proposed 1-glass RPC design, to be used in future applications, will avoid this problem, as the readout board will be an integral part of the chamber design (see below).

### 3.2 Further R&D

In preparation for a bid to design and build the hadron calorimeter of a detector at a future Linear Collider, the DHCAL group is investigating a number of R&D topics, such as a gas recirculation system, a high voltage distribution system, improved front-end electronics with token ring passing and the ability to pulse the power, etc. However, of particular interest is the development of a new design for the active elements, i.e. the RPCs. The new design features only a single resistive plate (in this case glass) and places the signal pads directly into the gas volume. This new design offers a number of distinct advantages: an average pad multiplicity close to unity and almost independent of the efficiency (simplifying the equalization procedure), a thinner chamber design (important for calorimetry), a higher rate capability (required in the forward direction), and a better position resolution (the exact improvement is still to be determined.)

To date, several small ($20 \times 20$ cm$^2$) and large ($32 \times 48$ cm$^2$) chambers have been built based on this new design, see Figure 8. Their performance was tested with cosmic rays and proved to be similar to the performance of the 2-glass design, but with an average pad multiplicity close to one. Tests in the Fermilab test beam are planned for the near future.

### 4. Conclusions

In the context of detector R&D for a future Linear Collider, an imaging hadron calorimeter based on Resistive Plate Chambers is being developed, the DHCAL. The concept was validated with the construction and thorough testing of a large-scale prototype in the Fermilab and CERN test beams. A number of (minor) operational issues were discovered during the prolonged test beam activity, which can be remedied with simple design changes. The DHCAL group is pursuing further R&D, in particular the validation of a novel 1-glass RPC design, in preparation for a bid to design and build the hadron calorimeter of a future Linear Collider experiment.

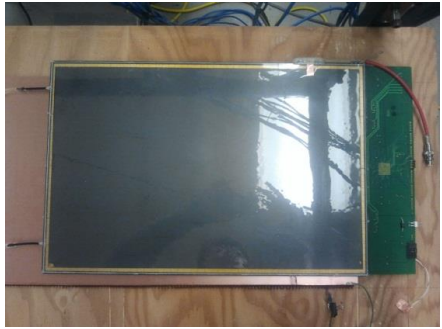

**Figure 8.** Photograph of a 1-glass RPC. The readout pads can be seen through the resistive paint of the glass plate.

## Acknowledgments

The author would like to thank the organizer, in particular Professor Yi Wang, for the warm hospitality and the well organizer meeting.

## References

[1] M.A. Thomson, *Particle flow calorimetry and PandoraPFA algorithm*, Nucl. Instrum. Meth. **A611**:25-40, 2009.

[2] G. Drake et al., Resistive *Plate Chambers for Hadron Calorimetry: Tests with Analog Readout*, Nucl. Instr. and Meth. A578, 88 (2007)

[3] https://twiki.cern.ch/twiki/bin/view/CALICE/WebHome

[4] A. Bambaugh et al., *Production and commissioning of a large prototype Digital Hadron Calorimeter for future colliding beam experiments,* Nuclear Science Symposium and Medical Imaging Conference (NSS/MIC), 2011 IEEE; DOI: 10.1109/NSSMIC.2011.6154437

[5] C. Adloff et al., *Construction and Commissioning of the CALICE Analog Hadron Calorimeter Prototype*, 2010 JINST 5 P05004.

[6] B.Bilki et al., *Hadron showers in a digital hadron calorimeter*, 2009 JINST 4 P10008.